\documentstyle[aps,preprint]{revtex}

\begin{document}
\draft
\begin{center}
{\Large{\bf Electromagnetic self-energies of the neutral pions:  
Contributions from vector resonances}}\\[4mm]
Dao-Neng Gao\\
{\small 
Center for Fundamental Physics,
University of Science and Technology of China\\
Hefei, Anhui, 230026, People's Republic of China
\\
and\\
International Centre for Theoretical Physics,
P.O.Box 586, 34100 Trieste, Italy
}\\[2mm]
Mu-Lin Yan\\
{\small China Center of Advanced Science and Technology (World Lab)\\
P.O.Box 8730, Beijing, 100080, People's Republic of China\\
and\\
Center for Fundamental Physics,
University of Science and Technology of China\\
Hefei, Anhui, 230026, People's Republic of China\footnote{Mailing
address}}
\end{center}

\begin{abstract}
\noindent
By employing vector meson dominance to establish the electromagnetic interactions of
hadrons, the electromagnetic self-energy of the neutral pions, which
receives the
contributions from the vector resonances due to chiral anomaly, is  
evaluated. 
This part of the electromagentic mass, which can contribute to the mass
difference of $\pi^\pm-\pi^0$, is finite and as expected, is smaller than the
contributions due to isospin symmetry breaking and electromagnetic chiral logs
corrections, however, could be compared with the contribution from the short
distance QCD.
\end{abstract}


In the limit of vanishing light quark masses, quantum chromodynamics (QCD)
lagrangian possesses an exact chiral SU(3)$_L\times$SU(3)$_R$ symmetry, which is
spontaneously broken
to SU(3)$_V$ vectorial symmetry with appearance of eight Goldstone bosons
($\pi$, $K$,and $\eta$).  At low energies, these Goldstone pseudoscalar particles play an
important role in chiral dynamics of the strong interaction. 

The mass difference between charged and neutral pions is one of the
fundamental properities of pion system which is accessible to the accurate
measurement. It is of interest to evaluate this quantity to increase the  
understanding of the low energy dynamics in hadronic physics.
It is known from chiral perturbation theory to one-loop \cite{CHPT} that the
contribution to the 
mass splitting of $\pi^\pm-\pi^0$ due to isospin symmetry breaking $m_u\not=m_d$ is
small, and the mass difference of pions is almost entirely 
electromagnetic in orgin. The mass difference of pions due to 
the virtual photon contribution has been investigated by many
authors \cite{das67,Dashen,Ecker,Urech,emmass,MSR98}. The well-known result of
$m_{\pi^\pm}-m_{\pi^0}$ in the soft-pion limit was
obtained by Das {\it et al.} \cite{das67} based on current algebra method.
The same result was
subsequently recovered by employing chiral effective lagrangian in many
literatures \cite{Ecker,Urech,emmass}. 

Chiral perturbation theory for virtual electromagnetic effects was first described 
at lowest order by Dashen \cite{Dashen}. Urech has systematically studied the
next-to-leading order terms \cite{Urech}.  In
the chiral limit where the light quark masses are set zero, the mass difference of 
pions receives the lowest order contribution from the low-energy
effective lagrangian of QCD in the presence of electromagnetic interaction, which is  
$O(e^2 p^0)$ \cite{Ecker,Urech,MSR98},
\begin{equation}
{\cal L}_{\rm eff}=e^2 C ~Tr ~Q U Q U^\dagger + ...,
\end{equation}  
where $U={\rm exp}(\sqrt{2}i \pi/f_\pi)$, $\pi=\pi^a
\lambda^a$ being the matrix of the
ordinary psedudoscalar
Goldstone fields and $f_\pi$ pion decay constant, and $Q={\rm
diag}[2/3, -1/3, -1/3]$ is charged matrix associated
with the electromagnetic couplings of the light quarks. From eq. (1),
by expanding $U$ to the quadratic terms of the pseudoscalar fields
\begin{equation}
{\cal L}_{\rm eff}=-2 e^2 C \frac{1}{f_\pi^2}\pi^+\pi^-+...,
\end{equation}
showing that, in the presence of electromagnetic interactions and in the chiral
limit, the charged pion
becomes massive, however, the neutral pion still remains massless
\begin{equation}
{(m_{\pi^\pm}^2-m_{\pi^0}^2)}_{\rm EM}={(m_{\pi^\pm}^2)}_{\rm EM}=\frac{2 e^2
C}{f_\pi^2},\;\;\;\;\; {(m_{\pi^0}^2)}_{\rm EM}=0,
\end{equation}
which is in agreement with the result obtained by Dashen \cite{Dashen}. 

The $C$ in eqs. (1)-(3) is the low-energy coupling constant which is used
to determine the electromagnetic mass difference of
pions at the lowest order. The investigations  by Das {\it et al.}
\cite{das67}
and by Ecker et al \cite{Ecker} have shown that this coupling constant is
saturated with the low-lying axial-vector and vector resonances. 

The Dashen's relation [eq. (3)] is only valid in the chiral limit. When the
calculation is beyond the lowest order, i.e. the light quark has a small mass,
the charged and neutral pions will become massive, especially, the mass
of the neutral pions will receieve the contributions from the effects of the virtual
photon.  The
mass difference between $\pi^\pm$ and $\pi^0$ has been evaluated beyond the chiral
limit. The contributions to the $\pi^\pm-\pi^0$ mass splitting due to
$m_u\not=m_d$ and the chiral logs electromagnetic corrections have been
obtained in
Refs. \cite{CHPT,LP73,MK91}, and they are small compared to the lowest order
contribution in eq. (3).

As is well known that the pion electromagnetic mass difference amplitudes
receive different contributions from the long and short distance QCD with the first
one dominating the calculation. In the present paper, it will be shown
below that, besides the
contributions due to isospin symmetry breaking and the chiral logs
electromagnetic
corrections, there exists the contribution from the vector resonances to
electromagnetic self-energies of the neutral pions, which can contribute
to the
mass difference of pions in the long distance QCD beyond the chiral limit.
We will find that this
part of the electromagnetic self-energies of $\pi^0$ is finite and small,
however, can be compared with the contribution
from the short distance QCD \cite{Natale}. 

It is easy to find that the vector resonances contributions to the electromagnetic
self-energies of the neutral pions only come  from the interaction
vertices in the
odd-parity effective lagrangian due to chiral anomaly \cite{WZW}. The general
$VV\pi$
interaction vertices ($V$ denotes vector meson) can be expressed as the following
phenomenological lagrangians 
\begin{eqnarray}
{\cal L}_{\omega\rho\pi}=g_{\omega\rho\pi}
\epsilon^{\mu\nu\alpha\beta}\rho^i_{\mu\nu}\omega_{\alpha\beta}\pi^i,\\
{\cal L}_{\phi\rho\pi}=g_{\phi\rho\pi}
\epsilon^{\mu\nu\alpha\beta}\rho^i_{\mu\nu}\phi_{\alpha\beta}\pi^i,
\end{eqnarray}
where $V_{\mu\nu}=\partial_\mu V_\nu-\partial_\nu V_\mu$, $V$ denotes $\rho$,
$\omega$, and $\phi$-meson respectively. $g_{\omega\rho\pi}$ and $g_{\phi\rho\pi}$
are the corresponding coupling constants.  Here we do not adopt any
specific models to derive the above interaction lagrangians in order to
make our analysis in this paper in a model independent way.

On the other hand, in order to evaluate the virtual photon contribution to
the electromagnetic self-energies of the mesons, we should introduce the
electromagnetic interaction into the phenomenological lagrangian. It is well known
that vector meson dominance (VMD) \cite{Sak69,KLZ} has been well established in
studying the
electromagnetic interactions of hadrons. VMD predates the standard model. It has not
been derived from the standard model, but nevertheless enjoys phenomennological
supports in describing hadronic electromagnetic interactions. As pointed
out by Sakurai
\cite{Sak69}
and by Kroll, Lee, and Zumino \cite{KLZ}, VMD could be written as follows 
\begin{equation}
{\cal L}_{\rm VMD}=\frac{em_\rho^2}{g_\rho}A^\mu
\rho^0_\mu+\frac{em_\omega^2}{g_\omega}A^\mu
\omega_\mu+\frac{em_\phi^2}{g_\phi}A^\mu \phi_\mu,
\end{equation}
where $g_\rho$, $g_\omega$, and $g_\phi$ are the coupling constants.  Thus combining
eqs. (4), (5) with (6), we can
calculate the $S$ matrix which will contribute to the electromagnetic self-energies
of the neutral pions:
\begin{equation}
S_\pi=\langle \pi|T~ {\rm exp}[i\int d^4 x ({\cal L}_{VV\pi}+{\cal L}_{\rm
VMD})]-1|\pi\rangle.
\end{equation}
On the other hand, $S_\pi$ can also be expressed in terms of the effective
lagrangian of the mesons as
\begin{equation}
S_\pi=\langle\pi|i\int d^4 x {\cal L}_{\rm eff}(\pi)|\pi\rangle.
\end{equation}
Noting that ${\cal L}_{\rm
eff}=\frac{1}{2}\partial_\mu\pi\partial^\mu
\pi-\frac{1}{2}m^2_\pi\pi^2$, Then the electromagnetic interaction correction to the
mass of the neutral pions reads
\begin{equation}
\delta m^2=\frac{2iS_\pi}{\langle\pi|\pi^2|\pi\rangle},
\end{equation}
where $\langle\pi|\pi^2|\pi\rangle=\langle\pi|\int d^4 x~ \pi^2(x)|\pi\rangle$.

At one loop level and $O(e^2)$, it is easy to obtain
\begin{equation}
S_\pi^\omega=\langle\pi|\frac{i^4}{4!} 6 \int d^4 x_1 d^4 x_2 d^4 x_3 d^4 x_4 ~T~
{\cal L}_{\omega\rho\pi}(x_1){\cal L}_{\omega\rho\pi}(x_2)
{\cal L}_{\rho\gamma}(x_3){\cal L}_{\rho\gamma}(x_4)|\pi\rangle,
\end{equation}
\begin{equation}
S_\pi^\phi=\langle\pi|\frac{i^4}{4!} 6 \int d^4 x_1 d^4 x_2 d^4 x_3 d^4 x_4 ~T~    
{\cal L}_{\phi\rho\pi}(x_1){\cal L}_{\phi\rho\pi}(x_2)
{\cal L}_{\rho\gamma}(x_3){\cal L}_{\rho\gamma}(x_4)|\pi\rangle,
\end{equation}
where ${\cal L}_{\rho\gamma}$ is the direct interaction vertex of $\rho^0$ and the
photon from VMD [eq.(6)]. Here we ingore $S_\pi^\rho$ which comes from
combining ${\cal L}_{VV\pi}$ with ${\cal L}_{\omega\gamma}$ or  
${\cal L}_{\phi\gamma}$ because this part of the contribution only makes 
the same
shift to the charged and neutral pions and no contribution to the mass difference of
$\pi^\pm-\pi^0$.

Thus from eqs. (4)-(6) and (9)-(11), we will get
\begin{equation}
(m_{\pi^0}^2)_{\rm EM}^\omega=-2ie^2 (\frac{4
g_{\omega\rho\pi}}{g_\rho})^2\int\frac{d^4 k}{(2\pi)^4}\frac{p^2 k^2
-(p\cdot k)^2}{k^2 [(k-p)^2-m_\omega^2](k^2-m_\rho^2)^2},
\end{equation}
\begin{equation}
(m_{\pi^0}^2)_{\rm EM}^\phi=-2ie^2 (\frac{4
g_{\phi\rho\pi}}{g_\rho})^2\int\frac{d^4 k}{(2\pi)^4}\frac{p^2 k^2
-(p\cdot k)^2}{k^2 [(k-p)^2-m_\phi^2](k^2-m_\rho^2)^2}.
\end{equation}
The integrations in the above equations are free of divergences, therefore
the calculations on them are straightforward,
\begin{eqnarray}
(m_{\pi^0}^2)_{\rm EM}^\omega=-\frac{3 e^2}{16\pi^2}(\frac{4
g_{\omega\rho\pi}}{g_\rho})^2m_\pi^2 m_\rho^2 \int^1_0 d x_1\int^{x_1}_0 d x_2
\frac{x_2}{x_2+\frac{m_\omega^2}{m_\rho^2}(1-x_1)-\frac{m_\pi^2}{m_\rho^2}x_1(1-x_1)},\\
 (m_{\pi^0}^2)_{\rm EM}^\phi=-\frac{3 e^2}{16\pi^2}(\frac{4
g_{\phi\rho\pi}}{g_\rho})^2m_\pi^2 m_\rho^2 \int^1_0 d x_1\int^{x_1}_0 d x_2
\frac{x_2}{x_2+\frac{m_\phi^2}{m_\rho^2}(1-x_1)-\frac{m_\pi^2}{m_\rho^2}x_1(1-x_1)},
\end{eqnarray}
\begin{equation}
(m_{\pi^0}^2)_{\rm EM}^V=(m_{\pi^0}^2)_{\rm EM}^\omega+(m_{\pi^0}^2)_{\rm
EM}^\phi.
\end{equation}

The unknown coupling constants in eqs. (14) and (15) are 
$g_{\omega\rho\pi}$, $g_{\phi\rho\pi}$, and $g_\rho$. By employing the
$\rho-\gamma$
vertex in eq. (5), the decay width of $\rho\rightarrow e^+e^-$ is 
\begin{equation}
\Gamma(\rho\rightarrow e^+e^-)=\frac{4\pi\alpha_{\rm
EM}^2}{g_\rho^2}\frac{m_\rho}{3}.
\end{equation}
Here $\alpha_{\rm EM}=\frac{e^2}{4\pi}$. The experimental data is 6.77$\pm$0.32 keV 
\cite{PDG98},
therefore one can determine $g_\rho$ from eq. (17). As to
$g_{\omega\rho\pi}$ and $g_{\phi\rho\pi}$, however, things will be complicated. It
is not
straightforward to determine $g_{\omega\rho\pi}$ by employing the data of
$\Gamma(\omega\rightarrow 3 \pi)$ because there exists the unknown direct coupling
of the process $\omega\rightarrow 3 \pi$ besides
$\omega\rightarrow\rho\pi\rightarrow (2\pi)\pi$.
The data of $\Gamma(\phi\rightarrow\rho(2\pi)\pi)$ and
$\Gamma(\phi\rightarrow
3\pi)$ by the direct decay are provided separately before 1998
\cite{PDG96} while only the total width
of these two
channels is provided in Particle Data Booklet 1998 \cite{PDG98}. 

Fortunately, if watching eqs(14) and (15) carefully, one will find that we
only need the ratios of $\frac{g_{\omega\rho\pi}}{g_\rho}$ and
$\frac{g_{\phi\rho\pi}}{g_\rho}$, not the separate values of $g_{\omega\rho\pi}$,
$g_{\phi\rho\pi}$, and $g_\rho$, to evaluate the
electromagnetic self-energies of the neutral pions. Thus, we can
determine these two ratios directly by employing the experimental
information of the radiative decay processes $\omega\rightarrow\pi\gamma$
and $\phi\rightarrow\pi\gamma$.  From eqs. (4), (5), and (6), the decay widths of
these two channels read
\begin{equation}
\Gamma(\omega\rightarrow\pi\gamma)=\frac{2\alpha_{\rm
EM}}{3}(\frac{g_{\omega\rho\pi}}{g_\rho})^2 m_\omega^3
(1-\frac{m_\pi^2}{m_\omega^2})^3,
\end{equation}
\begin{equation}
\Gamma(\phi\rightarrow\pi\gamma)=\frac{2\alpha_{\rm
EM}}{3}(\frac{g_{\phi\rho\pi}}{g_\rho})^2 m_\phi^3
(1-\frac{m_\pi^2}{m_\phi^2})^3.
\end{equation}
Experimentally, the branch ratios of these two channels are
$B(\omega\rightarrow\pi\gamma)=(8.5\pm0.5)\%$, and
$B(\phi\rightarrow\pi\gamma)=(1.31\pm0.13)\times 10^{-3}$ \cite{PDG98}. So 
we
obtain
\begin{eqnarray}
|\frac{g_{\omega\rho\pi}}{g_\rho}|=0.581\pm 0.017~ {\rm GeV}^{-1},\\
|\frac{g_{\phi\rho\pi}}{g_\rho}|=0.0345\pm 0.0017~ {\rm GeV}^{-1},
\end{eqnarray}
\begin{equation}
|\frac{g_{\phi\rho\pi}}{g_{\omega\rho\pi}}|=0.0593.
\end{equation}
Then the corresponding numerical results of the electromagnetic
self-energies of the
neutral pions are
\begin{eqnarray}
(m_{\pi^0}^2)_{\rm EM}^\omega=-2.65\times 10^{-5}~ {\rm GeV}^2,\\
(m_{\pi^0}^2)_{\rm EM}^\phi=-7.65\times 10^{-8}~ {\rm GeV}^2.
\end{eqnarray}
Obviously, the contribution [eq. (24)] from the interaction ${\cal L}_{\phi\rho\pi}$
could be neglected. Therefore the neutral pion's mass shift
contributing to the mass difference of $\pi^\pm-\pi^0$ and receiving
the contributions from the vector resonances due to chiral anomaly is 
\begin{equation}
(\delta m_{\pi^0})^V_{\rm EM}=-0.096 ~{\rm MeV},
\end{equation} 
which is smaller than the contribution from the chiral logs electromagnetic
corrections \cite{MK91}, however, its absolute value is larger than the 
one from short distance QCD ($\Delta_{\rm SD}=-0.02$~MeV) in Ref.
\cite{Natale}.  

The $\phi$-meson contribution is negligiable because the ratio of $g_{\phi\rho\pi}$
and $g_{\omega\rho\pi}$ is very small [eq. (22)].  If the mixing of the $\omega$ and
$\phi$-mesons is ideal, $\omega$ is free of $s\bar{s}$, and $\phi$ is pure
$s\bar{s}$, the decay process $\phi\rightarrow \rho\pi$ is forbidden 
in the large $N_c$ limit. This is implied by the
Okubo-Zweig-Iizuka (OZI) rule \cite{OZI}. Corrections to the decay width
of this process come from
the deviation of the $\omega-\phi$ ideal mixing and the next leading order
contribution in the large $N_c$ expansion, 
which could be estimated in this paper. From eq. (5), it is easy to get
\begin{eqnarray}
\Gamma(\phi\rightarrow\rho\pi)=\frac{g_{\phi\rho\pi}^2}{6\pi m_\phi^3}
(m_\phi^2+m_\rho^2-m_\pi^2)(m_\phi^4+m_\rho^4+m_\pi^4-2m_\rho^2m_\phi^2-
2m_\phi^2 m_\pi^2-2m_\rho^2 m_\pi^2).
\end{eqnarray}

Using eqs. (17), (21), and the experimental value of $\Gamma(\rho\rightarrow
e^+e^-)$, one can estimate $|g_\rho|=5.04$ and $ |g_{\phi\rho\pi}|=0.174 ~{\rm
GeV}^{-1}$. The $\Gamma(\phi\rightarrow\rho\pi)$ is about 0.337 MeV, and  
the corresponding 
experimental value from Particle Data Booklet in 1996 is $0.570\pm 0.030$ MeV
\cite{PDG96}. 
Because the value of $g_{\phi\rho\pi}$ is determined from the decay
$\Gamma(\phi\rightarrow\pi\gamma)$, it seems  that there is some
inconsistency 
in connecting $\phi\rightarrow\pi\gamma$ with $\phi\rightarrow\rho\pi$ by VMD in
eqs. (5) and (6). This means that something is missing in constructing the
phenomenological interactions of ${\cal L}_{\phi\rho\pi}$ and ${\cal
L}_{\phi\pi\gamma}$ here. However,
for
$g_{\omega\rho\pi}$ is much larger than $g_{\phi\rho\pi}$ [even if one estimates the
latter one  by using eq. (26) instead of eq. (19)], the $\phi$-meson
contribution to electromagnetic
self-energies of the neutral pion is still negligible.

To conclude, by constructing the odd-parity effective lagrangian which is involved 
with the vector resonances and the neutral pions, and by employing VMD
to establish the electromagentic interaction of hadrons, we have
calculated the vector resonances contribution to the electromagnetic self-energies
of the neutral pions which can contribute to the mass difference of 
$\pi^\pm-\pi^0$. This part of the electromagentic mass is finite and
small compared to the other long distance contributions, however, could
be compared with the short distance QCD contribution.

\vspace{0.5cm}
One of the authors (D.N.Gao) wishes to acknowledge the International
Centre for Theoretical Physics, Trieste, Italy for its hospitality.
This work is partially supported by NSF of China through Chen Ning Yang.

\end{document}